\documentclass[paper,11pt]{JHEP}
\pdfoutput=1
\usepackage{graphicx}
\usepackage[centertags]{amsmath}
\usepackage{amsfonts}
\usepackage{amssymb} 
\usepackage{amsthm}
\usepackage{slashed}
\usepackage{enumerate,enumitem}
\usepackage{amsbsy}
\usepackage{url} 

\def\({\left(}
\def\){\right)}
\def\[{\left[}
\def\]{\right]}

\newcommand{\rr}{R}
\newcommand{\LL}{L}

\def\one{{\rm 1\kern -.9mm l}}                             %

\def\beq{\begin{equation}}
\def\eeq{\end{equation}}
\def\beqa{\begin{eqnarray}}
\def\eeqa{\end{eqnarray}}
\newcommand{\eqa}{\begin{eqnarray}}
\newcommand{\ena}{\end{eqnarray}}

\newcommand{\bz}{\bar{z}}
\DeclareMathAlphabet{\mathpzc}{OT1}{pzc}{m}{it}

\newcommand{\bT}{\bar{T}}
\newcommand{\R}{{\mathbb{R}}}

\newcommand{\Lag}{{\mathcal{L}}}
\newcommand\blank[1]{}

\newcommand{\fract}[2]{{\textstyle\frac{#1}{#2}}}

\newcommand{\eps}{\varepsilon}

\newcommand{\balpha}{\alpha\kern -6.7pt\alpha}
\newcommand{\bbalpha}{\alpha\kern -4.95pt\alpha}

\newcommand\en{\end{equation}}
\newcommand\bea{\begin{eqnarray}}
\newcommand\eea{\end{eqnarray}}
\newcommand\nn{\nonumber}

\newcommand\Epsilon{{\cal E}}

\newcommand{\One}{{\hbox{{\rm 1{\hbox to 1.5pt{\hss\rm1}}}}}}
\renewcommand{\One}{{\mathbb 1}}
\renewcommand{\One}{{\rm 1\!\!1}}

\newcommand{\ba}{\begin{eqnarray}}
\newcommand{\ea}{\end{eqnarray}}

\newcommand{\be}{\begin{equation}}
\newcommand{\ee}{\end{equation}}

\renewcommand{\log}{\ln}

\newcommand{\vg}{\vec{ \bf g } }
\newcommand{\kersg}{ \mathcal{ K } }

\newcommand{\CFT}{\text{\tiny CFT}}

\newcommand{\TbT}{\text{T}\bar{\text{T}}}
\newcommand{\p}{(+)}
\newcommand{\m}{(-)}

\renewcommand{\ell}{{\mathcal L}}

\setlist[itemize]{leftmargin=*}

\def\XXint#1#2#3{{\setbox0=\hbox{$#1{#2#3}{\int}$ }
\vcenter{\hbox{$#2#3$ }}\kern-.6\wd0}}

\title{\centerline{$\TbT$-deformed  2D  Quantum Field Theories}}
\author{ \centerline{Andrea Cavagli\`a$^{1}$, Stefano Negro$^{2}$, Istv\'an M. Sz\'ecs\'enyi$^{3}$,  Roberto Tateo$^{1}$  } \\
${}^{1}$\sl\small Dipartimento\ di Fisica
and INFN, Universit\`a di Torino, Via P.\
Giuria 1, 10125 Torino, Italy.
\\
${}^{2}$\sl\small LPTENS, Ecole Normale Sup\'erieure, PSL Research University, Sorbonne Universit\'es,
UPMC Univ. Paris 06, CNRS UMR 8549, 24 Rue Lhomond, 75005 Paris, France.
\\
${}^{3}$\sl\small Department of Mathematical Sciences, Durham University,
South Road, Durham DH1 3LE, United Kingdom. 
\\
\vspace{0.3cm}
\centerline{ \hspace{0.3cm}
\it cavaglia@to.infn.it, negro@lpt.ens.fr, i.m.szecsenyi@durham.ac.uk, tateo@to.infn.it}
}
\abstract{
It was noticed many years ago, in the framework of massless RG flows, that the irrelevant composite operator $\TbT$, built with the components of the energy-momentum tensor, enjoys very special properties in 2D quantum field theories, and can be regarded as a peculiar kind of integrable perturbation.  
Novel interesting features of this  operator have recently emerged from the study of effective string theory models.

In this paper we study further properties of this distinguished perturbation. We discuss how it affects the energy levels and one-point functions of a general 2D QFT in finite volume through a surprising relation with a simple hydrodynamic equation. In the case of the perturbation of CFTs, adapting a result by L\"uscher and Weisz we give a compact expression for the partition function on a finite-length cylinder and make a connection with the exact $g$-function method. We argue that, at the classical level, the deformation naturally maps the action of $N$ massless free bosons into the Nambu-Goto action in static gauge, in $N+2$ target space dimensions, and we briefly discuss a possible interpretation of this result in the context of effective string models. 
}
\begin{document}

\section{Introduction}

The study of integrable models has a long history scattered with many significant achievements, 
including a deeper understanding of phase transitions and other non-perturbative  properties of interacting statistical mechanical systems, the study of important low dimensional condensed matter models \cite{onsager1944crystal,baxter2007exactly,essler2005one}, and the  link with   gauge theories in two, three and four space-time dimensions  and with the  AdS/CFT correspondence \cite{Beisert:2010jr}. 

The discovery of  a connection between the exact S-matrix method, the Thermodynamic Bethe Ansatz (TBA) \cite{ZamolodchikovTBA} and the  effective string theory description of the 
chromoelectric flux tube in confining gauge theories is a  further surprising result recently obtained within the integrable model setup \cite{dubovsky2012solving,Dubovsky:2013gi}. 
In particular, it has been discovered  that the finite-size spectrum of a theory of free massless bosons modified by the inclusion of a very simple scalar scattering phase (CDD factor) can be computed exactly with the TBA method and that its analytic form is precisely the well-known expressions for the energy levels of the Nambu-Goto model obtained in the early days of string theory via covariant quantization \cite{goddard1973quantum}. 
This idea was readily adapted in \cite{Caselle:2013dra} to encompass the open string case, a framework more directly relevant to  the study of  
the  quark-anti-quark potential. In the same paper, the setup was generalised to describe the deformation, via the same CDD factor, of an arbitrary 2D conformal field theory (CFT).  A tight  link was observed to previous studies on massless RG flows  with leading IR attracting operator given by the $\TbT$ composite operator built with the chiral and anti-chiral components of the CFT stress-energy tensor (see \cite{zamolodchikov1991tricritical},\cite{zamolodchikov1994thermodynamics},\cite{klassen1992spectral}, \cite{dorey2000new}, \cite{dunning2002massless}). 
Earlier   observations on the special r\^ole played by the $\TbT$ operator in integrable perturbations of CFTs were recorded in a few occasions, mainly in the framework of Form Factors \cite{Fateev:1998xb, Delfino:2006te} for correlation functions, and hints on possible links with CDD factors and  TBA may be envisaged in \cite{Mussardo:1999aj}.
A general  quantum field theory definition of this special composite operator away from criticality was proposed by Sasha Zamolochikov in \cite{Zamolodchikov:2004ce}, a paper of central importance for  the  purposes of the current work.
In the context of relativistic integrable field theories with factorised scattering \cite{Zamolodchikov:1978xm}, the irrelevant perturbations considered in this paper are described by formally modifying the S-matrix with an extra diagonal phase shift, depending on a deformation parameter $t$\footnote{ In the effective string theory context, this parameter is related to the string tension $\sigma = \frac{1}{2 t}$.  }:
\beqa\label{eq:massiveCDD}
 S_{ij}^{kl}(\theta) \rightarrow S_{ij}^{kl}(\theta) \, e^{i \delta^{(t)}_{ij}(\theta) }  ,
\eeqa
where $\theta=\theta_i-\theta_j$ denote the relative particles' rapidity and 
\beqa\label{eq:genCDD}
\delta^{(t)}_{ij}(\theta) = \delta^{(t)}(m_i, m_j , \theta) =t \,   m_i m_j   \, \sinh(\theta ) .
\eeqa
When the masses of all physical particles are sent to zero, the net effect of the presence of the CDD factor is to introduce an interaction between left and right-moving massless excitations. The phase shift for this process is simply
\beqa\label{eq:NGCDD}
\delta_{ij}^{(t)}(\theta_i-\theta_j) = t \, \frac{M_i M_j }{2} \, e^{\theta_i-\theta_j}= -2 t \, p^{\p}_i\, p^{\m}_j ,
\eeqa
where $ p^{\p}_i$ and $ p^{\m}_j $ are the momenta of right and left moving particles, respectively, and $M_i$, $M_j$ fix the relative energy scales. 
The CDD factor (\ref{eq:genCDD}) was  introduced in  \cite{Dubovsky:2013ira} as a   generalization of  the proposal of \cite{dubovsky2012solving} to theories 
with more than one particle species. 
It was first pointed out in \cite{Caselle:2013dra} that this deformation corresponds to perturbing the CFT theory with a series of irrelevant operators whose leading term is $\TbT$. 
As we shall discuss in this paper, the introduction of the phase factor in the nonlinear integral equations describing the spectrum at finite-size leads to a characteristic deformation of the energy levels as $t$ is varied, which flows according to the inviscid Burgers equation of hydrodynamics. Further, the wave-breaking phenomenon, typical of the solutions to this simple fluid equation, is seen to correspond to the appearance of a spectral singularity at physical values of the system size: the famous tachyon singularity of effective string theories.  While the specific form of the deformation of the spectrum was obtained for integrable theories, the result is strongly connected to the general properties of the $\TbT$ operator studied in \cite{Zamolodchikov:2004ce}, suggesting that the latter results might have a broader range of validity. Indeed, we believe that the formula for the deformation of the energy levels is universal and valid also for
non-integrable theories deformed by the  $\TbT$ composite field\footnote{The definition based on CDD factors may appear to be only meaningful in an integrable context, however it has been pointed out in \cite{Dubovsky:2013ira} (see 
also \cite{Conkey:2016qju}) that a natural generalization of (\ref{eq:genCDD}) can be also defined in a theory with non-integrable scattering. It would be very interesting to find a more precise link between the work \cite{Dubovsky:2013ira} and the   general properties of $\TbT$ invoked in this paper and show the equivalence between the two approaches.}. 
However, without a precise characterization, at least at the level of an effective action, of the deformed theory, this might seem a void statement. 
It was put forward in \cite{Caselle:2013dra} that the full effective action might be obtainable by a universal method involving the recursive re-definition of the stress-energy tensor. 
In this paper, we show that this idea can be realized (although with some subtle variation). We demonstrate this by reconstructing the full Nambu-Goto action in D dimensions in static gauge starting from the action of $D-2$ free massless bosons; we also apply  the same method to single boson Lagrangians with generic potential. Next, we use an idea 
first proposed by L\"uscher and Weisz \cite{Luscher:2004ib} to reconstruct the  cylinder partition function of $t$-deformed rational CFTs, linking the result  to the exact g-function formula proposed for massless TBA flows in \cite{Dorey:2009vg}. 
 The effect of this deformation appears to be easily computable also for other physical observables; the last example we briefly discuss concerns one-point functions on a cylinder geometry. 

Finally, we would like to mention that, while we were working on this project, F. Smirnov and A. B. Zamolodchikov were independently addressing the question of integrability of $\TbT$-related perturbations using the form factor method \cite{AZTalk0,FSTalk} (see also  \cite{Lashkevich:2016gev} for a related work).
The current project was mainly motivated by the questions raised in \cite{Caselle:2013dra} in connection to the results of \cite{Zamolodchikov:2004ce}, and we had already obtained all the results on the spectrum, including the connection to hydrodynamic equations, before we became aware that the same observation was made in a seminar by Zamolodchikov \cite{AZTalk}. The results in Section \ref{sec:FreeBoson}  on the  recursive construction of the perturbed effective action, were instead triggered by the observation made in \cite{AZTalk} on the deformation of the action. 

\section{The NLIE for CFTs and its CDD-factor deformation}\label{subsec:DDV}

Although most of the results discussed in the following have a more general validity, for concreteness we shall illustrate them in a very simple class of models, corresponding to the sine-Gordon model, its quantum reductions \cite{Smirnov:1990vm,Bernard:1990cw} and their CFT limits. Finite-size effects in these theories are compactly described by a single nonlinear integral equation (NLIE) \cite{klumper1991central, KlumperPearce0, KlumperPearce, DDV, DDV2}. We will briefly review this formulation and then discuss the effects of the $\TbT$ deformation. 

\subsection{The NLIE for CFTs}\label{sec:masslessDDV}
Let us start by recalling the application of the NLIE approach in the context of 2D CFTs with  (effective) central charge $\leq 1$, developed in \cite{Bazhanov:1994ft,BLZ2}. The object of these works was the unified description of the infinite set of integrals of motion (IMs) of the CFT. It was shown that, for any state of the theory, this information is stored into a pair of counting functions, $f^{\p}$ and $f^{\m}$ (corresponding to the right- and left-moving sector, respectively), each determined by a nonlinear integral equation. 
We consider the theory defined on a cylinder of radius $\rr$. The pair of NLIEs is characterized by a coupling $\xi$ and a twist parameter $\alpha$, and reads as follows:
\beqa\label{eq:DDVCFT}
f^{(\pm)}(\theta) &=& \pm i\alpha -i \frac{M \rr}{2} \,  e^{\pm \theta} \\
& \mp& \int_{\mathcal{C}^{(\pm)}_1} d y \, \kersg(\theta - y) \, \log\left( 1 + e^{\mp f^{(\pm)}(\theta)} \right) \pm  \int_{\mathcal{C}^{(\pm)}_2} d y \, \kersg(\theta - y) \, \log\left( 1 + e^{\pm f^{(\pm)}(\theta)} \right), \nn
\eeqa
where $M$ sets the energy scale, and the kernel is related to the sine-Gordon soliton-soliton scattering amplitude:
\beqa
\kersg(\theta) &=& \frac{1}{2 \pi i } \partial_{\theta} \log S_{sG}(\theta) , \\
\log S_{sG}(\theta) &=& - i \int_{0}^{\infty} \frac{d k}{k} \sin( k \theta) \,\frac{ \sinh\left( \pi k (\xi - 1)/2 \right) }{ \cosh\left( \pi k/2 \right) \, \sinh\left( \pi k \xi/2 \right) } .
\eeqa
Local IMs are related to the solutions of the NLIE \cite{BLZ2}. In particular, energy and momentum are expressible in terms of the lowest-order IMs:
 \beqa
 E(\rr) =  I^{\p}(\rr) + I^{\m}(\rr) , \;\;\;\;\; P(\rr) =  I^{\p}(\rr) - I^{\m}(\rr) ,
 \eeqa
 where
\beqa\label{eq:LocalIMs}
I^{(\pm)}(\rr)  &=&   \frac{M}{2} \,\left[\int_{\mathcal{C}^{(\pm)}_1} \frac{d\theta}{2 \pi i} \, e^{\pm\theta} \, \log\left( 1 + e^{-f^{(\pm)}(\theta)} \right)- \int_{\mathcal{C}^{(\pm)}_{2}} \frac{d\theta }{2 \pi i}\, e^{\pm\theta} \,  \, \log\left( 1 + e^{f^{(\pm)}(\theta)} \right) \right] . \nn
\eeqa
The information on the state is encoded in the choice of the integration contours $\mathcal{C}^{(\pm)}_{1}$, $\mathcal{C}^{(\pm)}_{2}$; for the ground state, one may take them to be straight lines slightly displaced from the real axis: $\mathcal{C}^{(\pm)}_{1} = \mathbb{R} + i 0^{+} $, $\mathcal{C}^{(\pm)}_{2} = \mathbb{R} - i 0^+$. Equations describing excited states have the same form but in general with deformed contours encircling a number of singularities $\{\theta^{\pm}_i\}$   with  $\left(1+ e^{f_{\pm}(\theta_i^{\pm})}\right) =0 $, see \cite{Bazhanov:1996aq, DT, FioravantiDDV, Feverati:1998dt}.

 It is not difficult to check that $E(\rr)$ and $P(\rr)$ have exactly the form predicted by conformal invariance. In particular, one may use the so-called dilogarithm Lemma \cite{DDV2} to show that
 \beqa\label{eq:Ipm}
 I^{\p}(\rr) =  2 \pi \, ( n_0 - c_{\text{eff}} /24 )/\rr, \;\;\;\;\; I^{\m}(\rr) = 2 \pi \, ({\bar{n}}_0 - c_{\text{eff}} /24 )/\rr,
 \eeqa
 where $c_{\text{eff}}$ is the effective central charge, parametrized as 
 \beqa
 c_{\text{eff}} = 
 1 - \frac{6 \, \alpha^2 }{\pi^2} \,\frac{\xi}{\xi+1},
 \eeqa
 and $n_0, \; \bar{n}_0 \in \mathbb{Z}$ characterize the excited state level and come about from the monodromies of the dilogarithm. 

\subsection{The deformation}
 Let us now come to the deformation. It is natural to conjecture that the CDD factor (\ref{eq:NGCDD}) is  equivalent to introducing a coupling between the counting functions corresponding to left and right movers. This is encoded in a new integral kernel $\chi_{CDD}$: 
 \beqa\label{eq:chiCDD}
\chi_{CDD}(\theta- \theta') = \frac{1}{2 \pi } \partial_{\theta} \delta_{CDD}(p^{\p}(\theta), p^{\m}(\theta')) = t \frac{M^2}{4 \pi} \,e^{\theta - \theta'},
\eeqa
so that equations (\ref{eq:DDVCFT}) are replaced by the system of two coupled nonlinear integral equations:
\beqa
f^{(\pm)}(\theta) &=&  \pm i \alpha -i \frac{M}{2} \,  e^{\pm\theta} \, \rr \\
&& \mp \int_{\mathcal{C}^{(\pm)}_1} d y \, \kersg(\theta - y) \, \log\left( 1 + e^{\mp f^{(\pm)}(y)} \right) \pm \int_{\mathcal{C}^{(\pm)}_2} d y \, \kersg(\theta - y) \, \log\left( 1 + e^{\pm f^{(\pm)}(y)} \right) \nn\\
&&\mp \int_{\mathcal{C}^{(\mp)}_1} d y \, \chi_{CDD}(\theta - y) \, \log\left( 1 + e^{\pm f^{(\mp)}(y)} \right) \pm \int_{\mathcal{C}^{(\mp)}_2} d y \, \chi_{CDD}(\theta - y) \, \log\left( 1 + e^{\mp f^{(\mp)} (y)} \right). \nn
\eeqa
Plugging in (\ref{eq:chiCDD}), it is simple to show that these equations can be rewritten as
\beqa\label{eq:chiCDDrewr}
f^{(\pm)}(\theta) &=&  \pm i \alpha -i \frac{M}{2} \,  e^{\pm\theta} \, \left( \rr  + 2 t E^{(\mp)}(\rr,t) \right) \\
&& \mp \int_{\mathcal{C}^{(\pm)}_1} d y \, \kersg(\theta - y) \, \log\left( 1 + e^{\mp f^{(\pm)}(y)} \right) \pm \int_{\mathcal{C}^{(\pm)}_2} d y \, \kersg(\theta - y) \, \log\left( 1 + e^{\pm f^{(\pm)}(y)} \right)
, \nn
\eeqa
where $E^{(\pm)}(\rr,t)$ denote the canonical expressions for $I^{(\pm)}$, evaluated on the solutions of the deformed NLIE system:
\beqa
E^{(\pm)}(\rr,t) = \frac{M}{2}\left[\int_{\mathcal{C}^{(\pm)}_1} \frac{d\theta}{ 2 \pi i } \, e^{\pm\theta} \, \log\left( 1 + e^{-f^{(\pm)}(\theta)} \right)- \int_{\mathcal{C}^{(\pm)}_2} \frac{d\theta}{ 2 \pi i } \,  e^{\pm\theta}  \,  \, \log\left( 1 + e^{f^{(\pm)}(\theta)} \right) \right].
\eeqa
 Equations (\ref{eq:chiCDDrewr}) reveal that the deformation can be interpreted as a redefinition of the length-parameters appearing in the NLIEs, $\rr \rightarrow \rr + 2t E^{(\pm)}(\rr,t)$. Consistency with (\ref{eq:Ipm}) then yields the following conditions: 
\beqa\label{eq:consistencyNG}
E^{\p}(\rr,t) = 2 \pi \,\left( \frac{ n_0 - c_{\text{eff}}/24 }{ \rr + 2 t E^{\m}(\rr,t) } \right), \;\;\;\;\; 
E^{\m}(\rr,t) = 2 \pi \,\left(\frac{ \bar{n}_0 - c_{\text{eff}}/24 }{ \rr + 2t E^{\p}(\rr,t) } \right).
\eeqa
These are precisely the relations found in \cite{Caselle:2013dra} starting from (generic) TBA equations and imply that the energy levels have the form \cite{Dubovsky:2013gi,Caselle:2013dra}:
\beqa\label{eq:NGEP}
E(\rr,t) &=& E^{\p}(\rr,t) + E^{\m}(\rr,t) \nn\\
&=&- \frac{\rr}{2 t} + \sqrt{ \frac{\rr^2}{4 t^2 } + \frac{2 \pi }{t} \, \left(n_0 + \bar{n}_0 -\frac{c_{\text{eff}}}{12} \right) + \left(\frac{2 \pi (n_0 - \bar{n}_0 )}{\rr}\right)^2 },  \\
P(\rr) &=&  E^{\p}(\rr) - E^{\m}(\rr) = \frac{ 2 \pi ( n_0 - \bar{n}_0 )}{\rr} .
\eeqa
 As reviewed in the introduction, for $c_{\text{eff}} = D-2$ this coincides with the spectrum of the Nambu-Goto string in $D$-dimensional target space obtained through light-cone quantization (for more comments on this relation, see the Conclusions). \\
 Let us also briefly mention that there are other NLIEs describing integrable CFTs, as well as massless flows  between minimal models \cite{zamolodchikov1994thermodynamics,dorey2000new,dunning2002massless}. The analysis of this section could be repeated without essential modifications to study the $t$-deformation of these systems as well. 
 The purpose of the following Section \ref{sec:sGDDV} is to illustrate the generalization of these results to the case of a massive integrable QFT, the sine-Gordon model.

\section{Deforming the sine-Gordon model}\label{sec:sGDDV}

The sine-Gordon model  can be seen as a relevant perturbation of the CFT corresponding to a single massless boson. 
The integrals of motion of the model are encoded in the single counting function $f(\theta)$, solution to the following nonlinear integral equation \cite{DDV2}:
\beqa\label{eq:DDVsG}
f(\theta) &=& -i m \rr \sinh(\theta) + i \alpha \nn\\
&&- \int_{\mathcal{C}_1} d y \, \kersg(\theta - y) \, \log\left( 1 + e^{ -f(y)} \right) + \int_{\mathcal{C}_2} dy \, \kersg(\theta - y) \, \log\left( 1 + e^{f(y)} \right) ,
\eeqa
where the kernel $\kersg$ is the same defined in Section \ref{sec:masslessDDV}, $m$ denotes the soliton mass, $\rr$ is the radius of the cylinder on which the theory is quantized, and the twist parameter $\alpha$ selects the vacuum (e.g., see \cite{Zamolodchikov:1994uw}). 
Again, the integration contours formally encode the characteristics of the state under consideration; for the ground state,  $\mathcal{C}_{1} = \mathbb{R} + i 0^{+} $, $\mathcal{C}_{2} = \mathbb{R} - i 0^+$. 
Energy and momentum can be obtained from the counting function through the relations:
\beqa
E(\rr) &=&  m \, \left[ \int_{\mathcal{C}_1} \frac{dy}{2 \pi i} \sinh(y)  \log\left( 1 + e^{-f(y)} \right)-\int_{\mathcal{C}_2} \frac{dy}{2 \pi i} \sinh(y)  \log\left( 1 + e^{f(y)} \right)  \right], \label{eq:EE}\\
P(\rr) &=& m \, \left[ \int_{\mathcal{C}_1} \frac{dy}{2 \pi i} \cosh(y)  \log\left( 1 + e^{ -f(y)} \right)-\int_{\mathcal{C}_2} \frac{dy}{2 \pi i} \cosh(y)  \log\left( 1 + e^{f(y)} \right)  \right] .\label{eq:PP}
\eeqa
In the case of two particles with equal mass, the CDD phase in (\ref{eq:massiveCDD}) takes the simple form
\beqa
\delta_{CDD}( \theta_1, \theta_2 ) = t\, m^2 \sinh( \theta_1 - \theta_2 ). 
\eeqa
This prompts us to deform the kernel appearing in the NLIE by
\beqa
\kersg(\theta) \rightarrow \kersg(\theta) + \frac{1}{2 \pi} \partial_{\theta} \delta_{CDD}(\theta) = \kersg(\theta) + t \frac{m^2}{2 \pi} \cosh(\theta) .
\eeqa
Inserting this new kernel in (\ref{eq:DDVsG}), after simple manipulations we find the deformed version of the NLIE:
\beqa\label{eq:modsGDDV}
f(\theta) &=& -i\, m\, \sinh(\theta) \left[ \rr + t \,E(\rr, t) \right] - i\, m\, \cosh(\theta) \,  t \, P(\rr, t)   \\ &&- \int_{\mathcal{C}_1} d y \, \kersg(\theta - y) \, \log\left( 1 + e^{-f(y)} \right) + \int_{\mathcal{C}_2} dy \, \kersg(\theta - y) \, \log\left( 1 + e^{f(y)} \right) ,\nn
\eeqa
where $E(\rr, t)$ and $P(\rr,t)$ are defined by the rhs of (\ref{eq:EE}),(\ref{eq:PP}) in terms of the solution to (\ref{eq:modsGDDV}).

\section{Deformation of the energy levels and hydrodynamic equations}\label{sec:FuncRels}

In this section, we describe the exact form of the energy levels in the presence of the CDD factor, 
showing that their deformation is ruled by a hydrodynamic equation. We will obtain this result  starting from the deformed NLIE for the sine-Gordon model (\ref{eq:modsGDDV}), but the derivation can easily be adapted to other integrable models and to other frameworks such as the TBA. 
 As a first preliminary observation, we point out that the quantization rule for the momentum does not change with the deformation. We expect that, for every state of the theory, the momentum will be quantized as
 \beqa\label{eq:quantP}
P(\rr, t) = P(\rr) &=& \frac{2 \pi k}{\rr}, \;\;\;\; k \in \mathbb{Z}, 
 \eeqa
where the integer $k$ is the same as for the undeformed solution, and depends on the topology of the integration contour\footnote{  It is not difficult to prove (\ref{eq:quantP}) using the standard dilogarithm trick, see for example the Lemma in Section 7 of \cite{DDV2}.}. 
The $k=0$ case is simpler and we shall treat it separately. 
\paragraph{Zero-momentum case: }
Equation  (\ref{eq:modsGDDV}) shows that, when $k=0$, the effect of the deformation can be regarded as an energy-dependent redefinition of the length,
\beqa
 f(\theta \,| \rr \,, \, t) = f(\theta \, | \,(\rr + t E(\rr , t)  )\, , 0 ) .
\eeqa
This observation implies a very simple relation between the energy, $E(\rr, t)$, computed in the presence of the deformation, and the undeformed energy:
\beqa\label{eq:functionalm}
E( \rr \,,\, t) = E((\rr + t E(\rr , t)) \, , 0 ) ,
\eeqa
which allows to compute the exact form of the $t$-deformed energy level once its $\rr$-dependence is known at $t=0$. 
It can be recognised that (\ref{eq:functionalm}) is precisely the implicit form of a solution of a well-known hydrodynamic equation:
\beqa\label{eq:invBurg}
\partial_{t} E(\rr,t) = E(\rr,t) \, \partial_{\rr} E(\rr,t) , 
\eeqa
where the deformation parameter plays the role of ``time'' variable, and the undeformed energy level serves as  initial condition at $t=0$. 

Equation (\ref{eq:invBurg}) describes an incompressible fluid in the absence of viscosity and pressure \cite{whitham2011linear}. It is often denoted as the inviscid Burgers equation. 
A crucial and well known aspect of the model is that its solutions tend to develop shock singularities, namely points where the gradient $\partial_{\rr} E$ blows up. As we will discuss shortly, they correspond to spectral singularities that may signal the breakdown of the theory at short distances, such as the famous Hagedorn singularity of the Nambu-Goto spectrum (see \cite{dubovsky2012solving,Caselle:2013dra} for a recent discussion of this phenomenon in the framework of TBA). 

\paragraph{General case: }
As observed in \cite{AZTalk}, a slightly more general hydrodynamic equation governs states with nonzero spin, $P(R) = 2 \pi k / \rr$, $k \neq 0$. To derive this result within the NLIE formalism, we start by rewriting equation (\ref{eq:modsGDDV}) as
\beqa\label{eq:modsGDDV2}
f(\theta) &=& -i \, m \, \mathcal{R}_0 \,  \sinh\left(\theta + \theta_0 \right)  + i \alpha \\ &&- \int_{\mathcal{C}_1} d y \, \kersg(\theta - y) \, \log\left( 1 + e^{-f(y)} \right) + \int_{\mathcal{C}_2} dy \, \kersg(\theta - y) \, \log\left( 1 + e^{f(y)} \right) ,\nn
\eeqa 
where the new parameters $\mathcal{R}_0 = \mathcal{R}_0\left(t,  \rr , E(\rr , t) \, \right) $, $\theta_0=\theta_0(t, \rr , E(\rr , t)) \,$ are defined through:
\beqa\label{eq:defineRtheta}
\, \mathcal{R}_0 \,
 \,
 \cosh\left( \theta_0
 \right)  &=&  R + t E(\rr , t)    , \\
 \mathcal{R}_0 \,
 \, \sinh\left(  \theta_0
 \right) &=&  t P(\rr)  .
\eeqa
 Equation (\ref{eq:modsGDDV2}) suggests that the solutions of the NLIE are still modified simply by a redefinition of the length and by a rapidity shift:
 \beqa\label{eq:fsigma}
 f(\theta | \rr , t ) = f(\theta + \theta_0 | \mathcal{R}_0 , 0 ).
 \eeqa 
 We can now use plug (\ref{eq:fsigma}) into (\ref{eq:EE}) to express the deformed energy $E(\rr \,  , t)$ in terms of quantities at $t=0$. After a change of variable in the integral, we obtain 
\beqa\label{eq:Funct}
 E(\rr \,, \, t ) &=& \cosh(\theta_0 ) \, E( \mathcal{R}_0 , 0 ) - \sinh(\theta_0 ) \, P(\mathcal{R}_0) ,
\eeqa
where $E( \mathcal{R}_0 , 0 )$ and $P( \mathcal{R}_0 )$ are energy and momentum evaluated in the unperturbed model at lengthscale $\mathcal{R}_0$. 
\beqa
E(\rr \, , \, t )  &=& \frac{1}{\mathcal{R}_0} \, \left[ \left(\rr + t E(\rr \, , \, t) \right) \, E(\mathcal{R}_0 , 0 ) - t P(\rr) \, P(\mathcal{R}_0)  \right] \\
&=& \frac{1}{\mathcal{R}_0} \, \left[ \left(\rr + t E(\rr \, , \, t) \right) \, E(\mathcal{R}_0 , 0 ) - t \frac{ 4 \pi^2 \, k^2 }{ \rr \, \mathcal{R}_0 } \right]. 
\label{eq:inSol}
\eeqa
It can be checked explicitly that (\ref{eq:inSol}) 
is an implicit form of the solution of the inviscid Burgers equation with a source term:
\beqa\label{eq:inhomBurg}
\partial_{t} E(\rr,t) = E(\rr,t) \, \partial_\rr E(\rr,t)  + \fract{P(\rr)^2}{\rr} ,  
\eeqa
where again the undeformed energy $E(\rr, 0)$ plays the role of initial condition at $t= 0$. 

Although our derivation was based on the NLIE, we expect that the hydrodynamic equations (\ref{eq:invBurg}) and (\ref{eq:inhomBurg}) (which were also found in \cite{AZTalk}) describe a uniquely-defined deformation of a generic 2D quantum field theory, generated by the operator $\TbT$. This is strongly suggested by the fact that the general properties of $\TbT$ established in \cite{Zamolodchikov:2004ce} hold for a generic 2D local quantum field theory, integrable or not. 

\subsection{Shock singularities and the Hagedorn transition}\label{sec:Hagedorn}
 To understand the occurrence of spectral singularities, let us briefly review a few well-known facts on the wave-breaking phenomenon in the inviscid Burgers equation, restricting to the $P=0$ case. 
For a generic initial condition, $E(\rr, 0)$, the model can be solved implicitly as
\beqa\label{eq:ERt}
E(\rr, t) = E( \tilde{\rr}(\rr, t),0 ),
\eeqa
with $\tilde {\rr}(\rr, t) = \rr + t \, E(\rr,t)$.  It can be shown (see, \cite{bessis1984pole}), that, at any fixed time $t > 0$,  the map $\rr \rightarrow \tilde{\rr}(\rr, t) $ has in general a  number of square-root branch points in the complex $R$-plane. 
To find their location (which depends on $t$)\footnote{For simplicity of notation, we will simply write $\rr_c$, $\tilde \rr_c$ for the position of a singularity in the $\rr$ and $\tilde \rr$-plane. However notice that these quantities depend  on $t$.}, it is convenient to consider the inverse map $ \tilde \rr \rightarrow  \rr( \tilde \rr,t ) = \tilde \rr  - t \, E( \tilde \rr,0 )$. A singularity is characterised by the condition 
\beqa\label{eq:R0c}
 \partial_{\tilde \rr} \, \rr( \tilde \rr \, , \, t )|_{\tilde \rr =\tilde \rr_c} = 1 - t \,\partial_{\tilde \rr} \, \left.  E( \tilde \rr,0 ) \right|_{\tilde \rr = \tilde \rr_{c} }= 0 .
\eeqa
Indeed, around a solution of equation (\ref{eq:R0c}), we can expand:
\beqa
 \rr( \tilde \rr,t )  \sim \rr_c + \mathcal{O}\left( \tilde \rr - \tilde \rr_c \right)^2 , \;\;\;\; \text{ for } \tilde \rr \sim \tilde\rr_c,
\eeqa
where $\rr_c \equiv  \rr( \tilde \rr_{c},t ) = \tilde \rr_{c} -  t \, E( \tilde\rr_{c},0 )$, and correspondingly,
\beqa\label{eq:branchcut}
\tilde{\rr}(\rr, t) \sim \tilde \rr_c + \mathcal{O}\left( \rr - \rr_c \right)^{\frac{1}{2} } , \;\;\;\; \text{ for }  \rr \sim \rr_c .
\eeqa
Relations  (\ref{eq:ERt}),(\ref{eq:branchcut}) imply that, for any solution $\tilde \rr_{c}$ of (\ref{eq:R0c}), one finds a singularity in the solution at $\rr = \rr_c$, which is generically a square-root branch point: $E(\rr, t) = E( \tilde \rr_c, 0 ) + \mathcal{O} ( R - R_c )^{\frac{1}{2} }$.

 In typical hydrodynamic applications, the initial profile is smooth on the real-$\rr$ axis, and for short times all branch points lie in the complex plane. The time evolution however in general brings one of the singularities on the real domain in a finite time, producing a shock in the physical solution. 
 
\begin{figure}[t]
\begin{minipage}{0.45 \textwidth}
 \centering
  \includegraphics[scale=0.6]{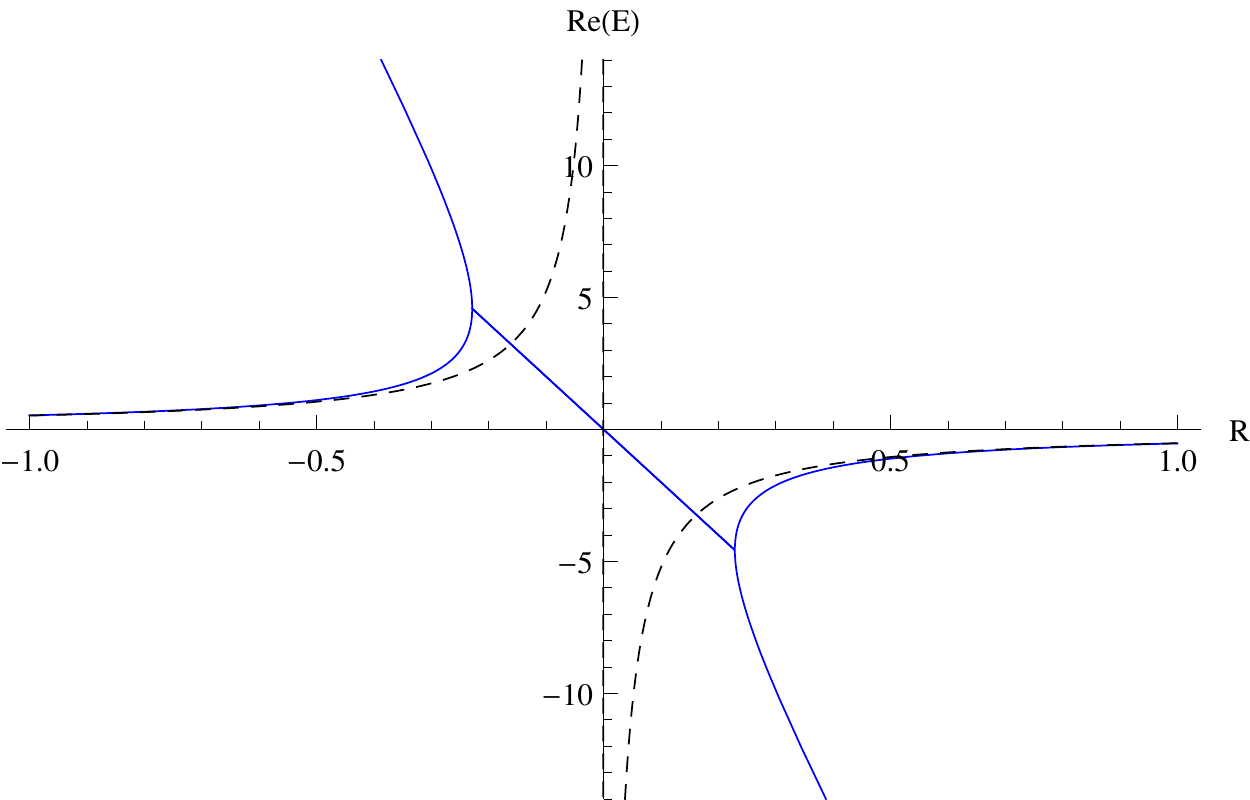}
  \caption{ Real part of $E(R, t)$ for $t=0$ (dashed line) and $t=0.025$ (solid line), for $c_{\text{eff}}=1$ \label{fig:NG} }
\end{minipage}
\hspace{1 cm}
\begin{minipage}{0.45\textwidth}
 \centering
  \includegraphics[scale=0.6]{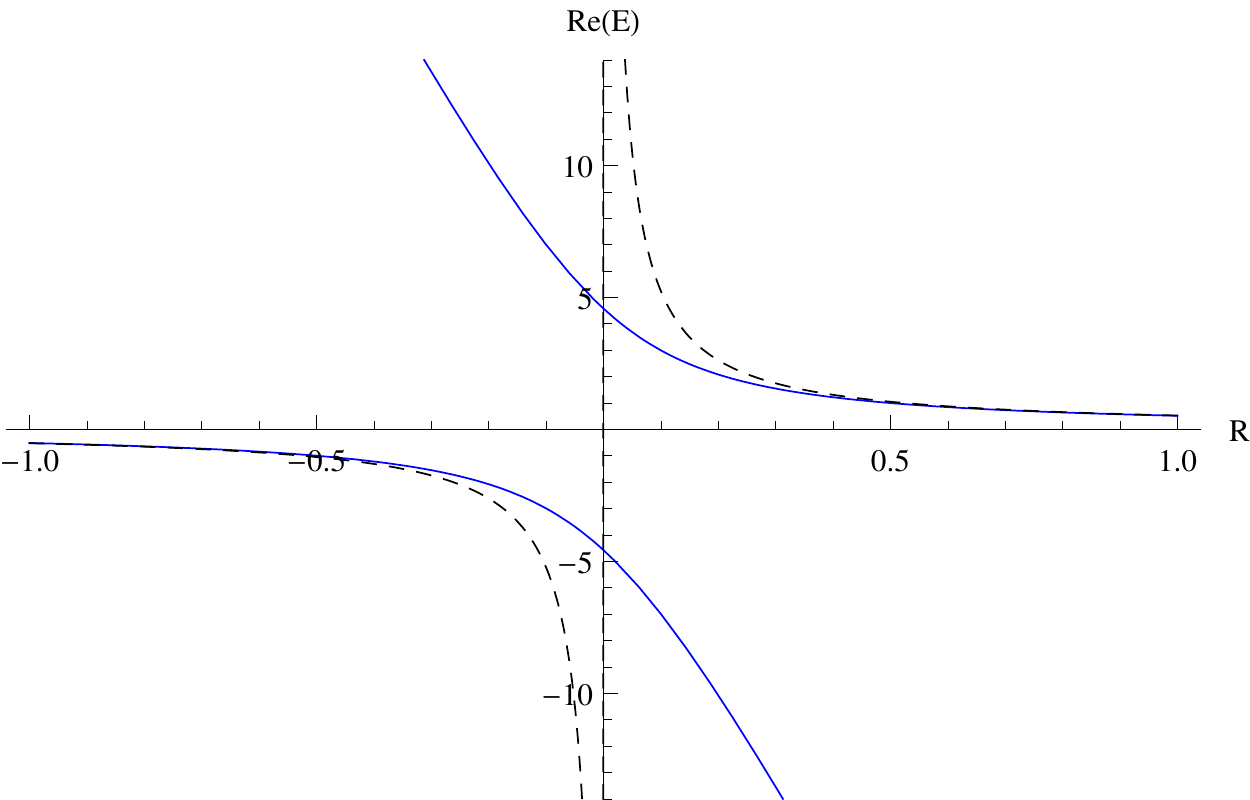}
    \caption{  Real part of $E(R, t)$ for $t=0$ (dashed line) and $t=0.025$ (solid line), for $c_{\text{eff}}=-1$ \label{fig:NG2} }
\end{minipage}
\end{figure}

Let us now turn to the typical situation one would encounter for the energy levels of a (UV-complete) QFT. Here, the ground state energy displays a pole at $\rr=0$, 
\beqa\label{eq:CFTpole}
E( \rr , 0 ) \sim - \pi \frac{ c_{\text{eff}}}{6 \, \rr} , \;\;\; \rr \sim 0 ,
\eeqa
where $c_{\text{eff}}$ is the effective central charge of the UV CFT. This behaviour implies that, for small times $t > 0$, equation (\ref{eq:R0c}) has two solutions very close to  $\widetilde \rr = 0$, satisfying
\beqa
(\widetilde \rr_c )^2 = \pi \frac{ c_{\text{eff}}}{6 } \, t + \mathcal{O}( t^2 ) ,
\eeqa
and correspondingly the solution is singular at
\beqa
\rr_c = \tilde \rr_c - t \, E( \tilde \rr_c , 0 ) \sim  t^{\frac{1}{2} } \, \sqrt{ \frac{2 \, \pi \, c_{\text{eff}}}{3}} + \mathcal{O}(t^{\frac{3}{2}}) .
\eeqa
In other words, as soon as $t > 0$, the pole at $\rr=0$ resolves into a pair of branch points. For the vacuum states with $c_{\text{eff}} > 0$, one of the branch points moves rightwards on the positive real-$\rr$ axis, producing a singularity for physical values of the radius as soon as $t > 0$. This is the above mentioned Hagedorn singularity (see Figure \ref{fig:NG} for an illustration). For states with $ c_{\text{eff}} \leq 0 $, instead, the branch points move off along the imaginary axis and there is no singularity for physical values of $\rr$ (see Figure \ref{fig:NG2}). 

All the features we have described are visible very clearly when the deformed theory is a pure CFT at $t=0$, as can be seen from the explicit form of the energy levels (\ref{eq:NGEP}). In particular, in that case the position of the singularities is exact:
 $\rr_c = \pm \sqrt{ \frac{2 \pi \, t \, c_{\text{eff}}  }{3} } $. 

As a last comment, one may wonder whether, for  a generic model, the flow could produce more exotic types of singularities occurring at later times, not originating from the pole at $\rr=0$. However, while in general a solution of the hydrodynamic equation will undergo a sequence of wave-breaking events (depending on the number of bumps in its initial profile), we found no evidence that these should occur in the physical region $\rr > 0$.

\section{Identification of the perturbing operator}
Let us now establish a direct link between the CDD-factor introduced in (\ref{eq:chiCDD}) and the $\TbT$ operator. 
We start from equation  (\ref{eq:inhomBurg}):
\beqa\label{eq:inhomBurg1}
\partial_{t} E_n(R,t)=   E_n(R,t) \partial_\rr E_n(R,t) + \fract{1}{\rr} P_n(R)^2  ,
\eeqa
where $E_n(R,t)$ denotes the $n$-th excited state of the system on a ring of size $R$. 
It is convenient to consider the theory as defined on a cylinder or torus, with Euclidean coordinates $(x, y) \sim (x+R, y)$. The expectation values of the components of the stress-energy tensor on the eigenstates of the Hamiltonian defined on constant-$y$ slices, satisfy \cite{ZamolodchikovTBA},
\beqa
E_n(R,t) = - R  \langle n| T_{yy}|n \rangle ,\,\,\,\, \partial_R E_n(R,t) = -\langle n| T_{xx}|n \rangle, 
\,\, \,\,  P_n= -i R \langle n| T_{xy}|n \rangle ,
\eeqa
so that (\ref{eq:inhomBurg1}) can be rewritten as
\beqa\label{eq:inhomBurg2}
\partial_{t} E_n(R,t) &=&  R \left(\langle n| T_{yy}|n \rangle \langle n| T_{xx}|n \rangle -   \langle n| T_{xy}|n \rangle  \langle n| T_{xy}|n \rangle \right) \nn \\  
                      &=&  -\fract{R}{\pi^2}  \left( \langle n| T|n \rangle \langle n| \bT|n \rangle - \langle n| \Theta|n \rangle  \langle n| \Theta|n \rangle \right).
\eeqa
In (\ref{eq:inhomBurg2}), we have used 
the standard conversion between the energy-momentum tensor components in Euclidean and complex coordinates,
\beqa
T_{xx}=- \fract{1}{2\pi} (\bT+T - 2 \Theta)\,,\,\, T_{yy}=  \fract{1}{2\pi} (\bT+T + 2 \Theta)\,,\,\,\ T_{xy}= \fract{i}{2 \pi} (\bT -T) .
\eeqa
One of the main  results of \cite{Zamolodchikov:2004ce} is the important identity
\beqa\label{eq:zam}
\langle n| \TbT|n \rangle  = \langle n| T|n \rangle \langle n| \bT|n \rangle - \langle n| \Theta|n \rangle  \langle n| \Theta|n \rangle,  
\eeqa 
which was established for a generic QFT in 2D and is based on a consistent, general definition, of  the $\TbT$ composite operator: 
\beqa\label{eq:TTbar}
\TbT(z,\bz) := \lim_{(z,\bz)  \rightarrow (z',\bz')} T(z,\bz)\bT(z',\bz') - \Theta(z,\bz) \Theta(z',\bz') + \text{total derivatives }.
\eeqa
Comparing (\ref{eq:zam}) with the rhs of (\ref{eq:inhomBurg2}) leads to the identification:
\beqa
\label{eq:pE}
\partial_{t} E_n(R,t) = - \fract{R}{\pi^2} \langle n| \TbT|n \rangle  = -\fract{1}{\pi^2} \langle n| \int_0^R dx \, \TbT(z,\bz) |n \rangle.
\eeqa
After multiplying by $L e^{ -L E_n(R,t)}$ and taking the trace, equation (\ref{eq:pE})  can be rewritten as
\beqa
\partial_{t}  \ln Z(R,L,t) =\fract{1}{\pi^2}\langle  \int_0^R dx \int_0^L dy \, \TbT(z,\bz) \rangle,      
\eeqa
where $Z(R,L,t)$ is the partition function of the perturbed system on a torus with characteristic lengths $(L,R)$. 
This suggests that, apart for total derivative terms, the Lagrangian density fulfils a simple local differential equation: 
\beqa
\partial_{t} \Lag(z, \bz,t)= -\frac{1}{\pi^2} \, \TbT(z, \bz,t) .
\label{eq:lagrangian}
\eeqa
We conclude that
\beqa
 \Lag(z, \bz,t) =  \Lag(z, \bz ,0) -\frac{t}{\pi^2} \, \TbT(z, \bar{ z } ) + \mathcal{O}\left(t^2\right) ,
\eeqa
generalizing the CFT result of \cite{Caselle:2013dra} to a massive quantum field theory.  Equation (\ref{eq:lagrangian}) first appeared in \cite{AZTalk} and is the main motivation for the analysis of the next Section.
\section{From the free boson to Nambu-Goto }\label{sec:FreeBoson}

\subsection{Single free boson}
The aim of this Section is to show how the perturbed action can be computed recursively from equation (\ref{eq:lagrangian}), and to verify that, starting from the free-boson theory, the deformation generates the Nambu-Goto action for a string moving in $3D$ target space written in  static gauge. 
We shall also present a closed formula for the infinite set of conserved currents, leading to local integrals of motion of this integrable theory. 
Our starting point is the free boson action :
\beqa
A_{\text{CFT}} = \frac{1}{4}  \int dx \, dy \, \left( (\partial_x\phi )^2 + (\partial_y\phi )^2 \right) = \frac{1}{2}  \int dz \,d\bar{z}  \, \left( \partial_z \phi \,  \partial_{\bar{z}} \phi \right) = \int dx \, dy \, \mathcal{L}_{\text{free}},
\eeqa
where $z = x+i y$, $\bar{z}=x-i y$, $\mathcal{L}_{\text{free}} = \partial \phi \, \bar \partial \phi$. 
For conciseness, let us introduce the rescaled energy-momentum components, $\tau = T/\pi$, $\bar{\tau}=\bar{T}/\pi$, $\theta=\bar \theta= \Theta/\pi$, and the composite operator $\tau  \bar{\tau }= \TbT/\pi^2 $, so that the differential equation (\ref{eq:lagrangian}) will take the form
\beqa\label{eq:lagrnew}
\partial_{t} \Lag(z, \bz,t)= -  \tau  \bar{\tau }(z, \bz,t) .
\eeqa
We will use the canonical expression for the energy-momentum tensor in a generic Lagrangian theory with a single boson field, which yields
\beqa\label{eq:Trule}
 \tau = - \frac{\partial\mathcal L}{\partial\(\bar\partial \phi \)}\partial\phi 
 ,\quad \bar \tau = - \frac{\partial\mathcal L}{\partial\(\partial \phi \)}\bar\partial\phi ,\quad
	\theta = \frac{1}{2} \,\left( \frac{\partial\mathcal L}{\partial\(\partial \phi \)}\partial\phi + \,\frac{\partial\mathcal L}{\partial\(\bar\partial \phi\)}\bar\partial\phi-  2  \mathcal L \right),
\eeqa 
so that (\ref{eq:lagrnew}) takes the form of a simple partial differential equation for the Lagrangian density
\beqa
\partial_{t} \Lag(z, \bz,t) = - \left( \frac{\partial\mathcal L}{\partial\(\bar\partial \phi \)}\partial\phi  \right)\, \left( \frac{\partial\mathcal L}{\partial\(\partial \phi \)}\bar\partial\phi\right) + \frac{1}{4} \left( \frac{\partial\mathcal L}{\partial\(\partial \phi \)}\partial\phi + \,\frac{\partial\mathcal L}{\partial\(\bar\partial \phi\)}\bar\partial\phi-  2  \mathcal L \right)^2 ,\label{eq:lagrdiff}
\eeqa
where we are taking into account the definition of the composite field $\tau \bar{\tau}$ as
\beqa
\tau\bar{\tau}(z,\bar{z}) = \tau(z ,\bar{z}) \, \bar{\tau}(z, \bar{z}) -\theta^2(z, \bar{z}).
\eeqa
In order to solve (\ref{eq:lagrnew}), we can setup a perturbative expansion in $t$,
\begin{equation}
	\Lag(z, \bz,t) = \sum_{j =0}^{\infty} t^j  \ell^{(j)} \;, 
\end{equation}
with initial condition $\ell^{(0)} \equiv \ell_{\text{free}}$, so that (\ref{eq:lagrdiff}) translates into
\begin{equation}
 \ell^{(j+1)}=-\frac{1}{j+1}\sum_{k=0}^j \left(\tau ^{(k)}\; \bar \tau^{(j-k)} - \theta^{(k)} \; \theta^{(j-k)} \right) ,
\label{eq:pertEQfor}
\end{equation}
with
\beqa
\tau^{(j)} = -  \frac{\partial\ell^{(j)}}{\partial\(\bar\partial \phi \)}\partial\phi 
 ,\quad \bar \tau^{(j)} = -\frac{\partial \ell^{(j)}}{\partial\(\partial \phi \)}\bar\partial\phi\;, \quad
	\theta^{(j)} = \frac{1}{2} \,\left( \frac{\partial  \ell^{(j)}}{\partial\(\partial \phi \)}\partial\phi + \,\frac{\partial\ell^{(j)}}{\partial\(\bar\partial \phi\)}\bar\partial\phi-  2 \ell^{(j)} \right)\; .\nn
\eeqa
The first order is the familiar result for the energy-momentum tensor of the free boson:
 \beqa
\tau^{(0)} = \tau_{\text{free}} =  - (\partial \phi )^2  \;\;\;\;\;\; \bar \tau^{(0)} =\bar \tau_{\text{free}} = -(\bar \partial \phi )^2 , \;\;\;\;\; \theta^{(0)} = \theta_{\text{free}} =0 .
 \eeqa
Going through a few more orders, it is rather simple to find the form of the general solution
\begin{equation}
	\ell^{(j)} = 
	\binom{1/2}{j+1}
       \; 2^{2j+1} \; \(\ell^{(0)}\)^{j+1}\; ,
\end{equation}
which can be verified by induction. Resumming the series, we find 
\beqa
\Lag(z, \bz,t)  = \frac{1}{ 2 t}\left( -1 + \sqrt{ 1 +4 t \,  \ell^{(0)} } \right) = -\frac{1}{2 t} + \mathcal{L}_{\text{NG}} \;.
\label{eq:def_lagr_oneBos}
\eeqa
The Lagrangian density $\mathcal{L}_{\text{NG}} $ appearing above is the Nambu-Goto Lagrangian for a bosonic string in $D=3$ target space:
\beqa\label{eq:genNG}
\mathcal{L}_{\text{NG}} \; dx dy=\frac{1}{2 t}\, \sqrt{\text{det}(\sum_{\mu=1}^{D}\partial_{\alpha} X^{\mu} \, \partial_{\beta} X^{\mu} ) }   \; dx dy, 
\eeqa
with $D=3$, in the so-called static gauge:
\beqa\label{eq:staticgauge}
X^1 \rightarrow x \;\;\;\; X^2 \rightarrow  y ,
\eeqa
where $x,\, y$ are Euclidean worldsheet coordinates, and where the transverse direction of oscillation $X^3$ is identified with the bosonic field, $ X^3 \rightarrow t^{\frac{1}{2}} \; \phi $. 

\paragraph{Integrals of motion:}
Let us conclude this Section by presenting the explicit form of the higher conserved currents of the model, $\left\{\tau_n, \; \bar \tau_n \right\}_{n \in \mathbb{N}}$, satisfying the conservation equations:
\beqa\label{eq:conservedNG}
{\bar \partial } { \tau }_n = {\partial} { \theta }_{n-2} , \;\;\;\;\;\; {\partial } { \bar \tau }_n = { \bar \partial} { \bar \theta }_{n-2} ,
\eeqa
which imply that the following quantities are local integrals of motion:
\beqa
I_{n-1} = \oint \left( \tau_n(z, \bar{z} ) \, dz + { \theta }_{n-2}(z, \bar z ) \, d{\bar z} \right) , \;\;\; \bar I_{n-1} = \oint \left( \bar \tau_n(z, \bar{z} ) \, d{\bar z} + {\bar  \theta }_{n-2}(z, \bar z ) \, dz \right),
\eeqa
where $I_1 + \bar I_1$ corresponds to the Hamiltonian of the theory. 
We obtained these currents as a deformation of following currents for the free theory at $t=0$:
\beqa
\tau^{(0)}_n = -({ \partial} \phi )^{n} , \;\;\; \bar  \tau^{(0)}_n = -({ \bar \partial} \phi )^{n} .
\eeqa
The strategy of the construction is simply to enforce perturbatively at every order in $t$, the compatibility between the conservation equation 
(\ref{eq:conservedNG}) and the equations of motion descending from the variation of the Lagrangian (\ref{eq:def_lagr_oneBos}), 
\begin{equation}
\sum_{n=0}^{\infty} (-1)^n    \left(\partial^n \fract{\partial {\cal L}}{\partial \partial^n \phi}   + \bar{\partial}^n \fract{\partial {\cal L}}{\partial \bar{\partial}^n \phi} \right)=0 ,
\end{equation}
which take the form
\begin{equation}
\partial \bar \partial  \phi= t \fract{ \bar{\partial}^2 \phi (\partial \phi)^2 +  \partial^2 \phi ( \bar{\partial} \phi)^2}{  1 + 2 t \partial \phi \, {\bar \partial} \phi}.
\end{equation}
Omitting the details of the calculation, the exact form of the currents is the following:
\beqa\label{eq:tautheta}
{ \tau }_n &=& -\frac{({ \partial} \phi )^{n} }{ \sqrt{1 + 4 t \; \mathcal{L}_{\text{free}} } } \, \left( \frac{ 2}{\sqrt{1 + 4 t \; \mathcal{L}_{\text{free}} }+1}\right)^{n-2}  , \\
 {  \theta }_{n-2} &=&  -t \,  \frac{ ({ \partial} \phi )^{n}  \,({ \bar \partial} \phi )^{2}  }{\sqrt{1 + 4 t \; \mathcal{L}_{\text{free}} } } \;
 \left( \frac{ 2}{\sqrt{1 + 4 t \; \mathcal{L}_{\text{free}} }+1}\right)^n ,
\eeqa
with a similar definition for $\bar \tau_n$ and $\bar \theta_{n-2}$.   
Notice that   ${  \tau }_2= \tau$ and   ${  \theta }_0= \theta$, where $\tau$ and $\theta$ are the components of the stress-energy tensor defined above, see (\ref{eq:Trule}). 

Finally, it is noteworthy that the following rather surprising relation between the trace of the stress-energy tensor and the composite field $\tau \bar{\tau}$ holds at all orders in $t$:
\beqa\label{eq:thetattb}
\theta(z,\bar{z})=-t\, \tau\bar{\tau}(z,\bar{z}) ,
\eeqa
which implies 
\beqa
\tau\bar{\tau}(z,\bar{z}) = \frac{1}{2 t^2} \, \left( \sqrt{1 + 4 t^2 \tau(z, \bar{z}) \, \bar{ \tau}(z, \bar{z}) } -1 \right).
\eeqa
\subsection{$N$ free bosons}
\label{subsec:Nfree}
Let us now consider the $t=0$ model as a theory of $N$ massless free bosons. It is simple to check directly that  the Lagrangian:
\begin{equation}
\mathcal L=-\frac{1}{2t}+  \mathcal L_{\textrm{NG}}^{N+2} , \,\, \mathcal L_{\textrm{NG}}^{N+2} =\frac{1}{2t} \, \sqrt{1 + 4 t\,  \mathcal L_{\textrm{free}}^N  - 4 t^2  \mathcal{B}_N}\; ,
\label{eq:Ddimboson_NG}
\end{equation}
with
\begin{equation}
\mathcal L_{\textrm{free}}^N =  \sum_{i=1}^N  \partial\phi_i\bar\partial\phi_i \;,\;\;\;\;
\mathcal B_N =\; \sum_{i=1}^N  \(\partial \phi_i\)^2 \, \sum_{j=1}^N \(\bar{\partial}\phi_j\)^2-\( \sum_{i=1}^N \partial\phi_i\bar\partial\phi_i\)^2\;,
\end{equation}
fulfils the appropriate generalization of (\ref{eq:lagrdiff}) to a system of $N$ bosonic fields. The Lagrangian $\mathcal L_{\textrm{NG}}^{N+2}$ defined above can again be derived from the Nambu-Goto action (\ref{eq:genNG}), now in $D=N+2$ dimensions, imposing the static-gauge condition (\ref{eq:staticgauge}) and identifying the fields as transverse oscillation modes of the string: $X^{j+2} = t^{\frac{1}{2}} \; \phi_j $, $j=1, \dots, N$. 
 One could also recover the same result from a perturbative construction similar to the one presented in the previous section. 
For completeness, we report the first few terms: setting $\mathcal L= \sum_{k=0}^{\infty} t^k \mathcal L^{(k)}$, we have
\beqa
\mathcal{L}^{(0)} &=& \mathcal L_{\textrm{free}}^N \,,\,\, \mathcal L^{(1)} = - \( \mathcal L_{\textrm{free}}^N \)^2- \mathcal B_N\,,\,\, \mathcal L^{(2)}= 2\(\mathcal L_{\textrm{free}}^N \)^3+2\mathcal L_{\textrm{free}}^N \,\mathcal B_N\;, \nn \\
\mathcal L^{(3)} &=&- 5\(\mathcal L_{\textrm{free}}^N \)^4-6\(\mathcal L_{\textrm{free}}^N \)^2 \, \mathcal B_N- \mathcal B_N^2\,, \,\,\dots
\label{eq:L_expans_NBos3}
\eea
\paragraph{Integrals of motion:}
An infinite set of integrals of motion can also be constructed for the $N>1$ case. The equations of motion in this case generalize to:
\begin{equation}
\partial \bar \partial  \phi_i   = t \;\left({ \bar{\partial}^2 \phi_i \;  \sum_{j=1}^N (\partial \phi_j )^2  +  \partial^2 \phi_i \;   \sum_{j=1}^N (\bar{\partial} \phi_j )^2  }\right)/{\left(1 + 2 t \;  \sum_{j=1}^N \partial \phi_j {\bar \partial} \phi_j \right)}.
\label{EOMN}
\end{equation}
Using (\ref{EOMN}), we proved that the following currents are conserved:
\beqa\label{eq:tauthetaN}
{ \tau }_n &=& -\frac{(-\tau_2^{(0)})^{n/2} }{ \sqrt{1 + 4 t \; \mathcal L_{\textrm{free}}^N  - 4 t^2 \;  \mathcal{B}_N  } } \, \left( 
\frac{1}{2} + t \;\mathcal L_{\textrm{free}}^N + \sqrt{1 + 4 t \; \mathcal L_{\textrm{free}}^N - 4 t^2 \; \mathcal{B}_N  } \right)^{1-n/2}  , \\
 {  \theta }_{n-2} &=& t \,  \frac{  (-\tau_2^{(0)} )^{n/2}  \,{ \bar{ \tau}_2^{(0)} }  }{\sqrt{1 + 4 t \; \mathcal L_{\textrm{free}}^N  - 4 \;t^2 \;  \mathcal{B}_N  } } \;
\left(\frac{1}{2} +t \;  \mathcal L_{\textrm{free}}^N  + \sqrt{1 + 4 t \;\mathcal L_{\textrm{free}}^N  - 4 t^2 \;  \mathcal{B}_N  } \right)^{-n/2},\eeqa
with
\beqa
\tau_2^{(0)} &=& -\sum_{j=1}^N (\partial \phi_j )^2 \,,\,\, \bar{\tau}_2^{(0)} = -\sum_{j=1}^N (\bar \partial \phi_j )^2  .
\eeqa
 Also in this case, there is a surprisingly simple relation between $\theta$ and $\tau\bar{\tau}$, given by (\ref{eq:thetattb}). 
\subsection{Single bosonic field with generic potential}
\label{subsec:FreeBosonPotential}
Finally we would like to consider another generalisation of the free boson, namely a system consisting of one boson interacting with an arbitrary potential $V(\phi)$:
\begin{equation}
	\mathcal L^V =  \, \partial\phi\bar\partial\phi + m^2 V[\phi]\;.
\end{equation}
Much in the same spirit as what we discussed for the free boson, we would like to follow the flow defined by the equation
\begin{equation}
\partial_{t} \mathcal L_t^V = -\tau \bar \tau\;,\qquad \mathcal L_{t=0}\equiv \mathcal L^{(0)} = \mathcal L^V .
\label{eq:diff_eq_again}
\end{equation}
Note that we are not requiring the potential $V[\phi]$ to give rise to an integrable theory. The only assumption we are making is that $V$ does not depend on the derivatives $\partial\phi$ and $\bar\partial\phi$. For $V[\phi] = \sin(\beta \phi)$, in particular, we recover the sine-Gordon model considered as our primary example in this paper -- however integrability does not play any special role in the following analysis. 
The first few terms in the perturbative expansion $\mathcal L^V_t = \sum_{j=0}^\infty t^j \ell^{(j)}\;$ are easily computed similarly to the previous sections:
\begin{align}
	\ell^{(1)} &= -\mathcal F^2 + \mathcal V^2 \;,\qquad \ell^{(2)}  = 2\mathcal F^3+\mathcal F^2\mathcal V + \mathcal V^3\;,\qquad \ell^{(3)}  = -5\mathcal F^4 - 4\mathcal F^3\mathcal V + \mathcal V^4\;,\nonumber\\
	\\
	\ell^{(4)} &= 14\mathcal F^5+15\mathcal F^4\mathcal V + 2\mathcal F^3\mathcal V^2+\mathcal V^5\;,\nonumber
\end{align}
where
\begin{equation}
	\mathcal F =  \partial\phi \bar\partial\phi\ \;,\qquad \mathcal V = m^2  V[\phi]\;.
\end{equation}
The general form of $\mathcal L^{(j)}$ is that of a polynomial in $\mathcal F$ and $\mathcal V$, whose coefficients are some combinatorial numbers. Although it is not immediately clear, it is easy to verify that the following closed form is valid: 
\beqa
	 \ell^{(j)}&=& \mathcal{V}^{j+1}  + (-1)^j \, \sum_{k=0}^{\lfloor \frac{j+1}{2} \rfloor } \frac{4^{j-k}}{\sqrt{\pi}} \frac{\Gamma\(j-k+\frac{1}{2}\)}{ ( {j-k+1}  )\Gamma\(j-2k+1\)}\,\frac{\mathcal F^{j-k+1}\mathcal V^k}{k!}\; \\
 &=&\mathcal V^{j+1}+\frac{(-4)^j}{\sqrt{\pi}}\frac{\Gamma\(j+\frac{1}{2}\)}{\Gamma\(j+2\)} \;\mathcal F^{j+1}\; {}_3F_2\(\left.\begin{array}{c c c c c}
	-1-j&,&\frac{1-j}{2}&,&-\frac{j}{2}\\
	&\frac{1}{2}-j&,&-j
	\end{array}\,\right\vert-\frac{\mathcal V}{\mathcal F}\)\;.\nonumber
\eeqa
The final expression for the $t \neq 0$ Lagrangian is
\begin{align}
	\mathcal L_t^V &= \frac{m^2V[\phi]}{1 - t m^2 V[\phi]}+\frac {1}{\sqrt{\pi}}\sum_{j=0}^\infty\(-4t\)^j \frac{\Gamma\(j+\frac{1}{2}\)}{\Gamma\(j+2\)} \, \(\,\partial\phi\bar\partial\phi\)^{j+1} 
	\nonumber \\
	& \;\;\times \,{}_3F_2\(\left.\begin{array}{c c c c c}
	-1-j&,&\frac{1-j}{2}&,&-\frac{j}{2}\\
	&\frac{1}{2}-j&,&-j
	\end{array}\,\right\vert-m^2\frac{V\[\phi\]}{\partial\phi\bar\partial\phi}\)\; ,\label{eq:VLagrangian}
\end{align}
which we did not manage to simplify further. This expression reduces to the Nambu-Goto result (\ref{eq:def_lagr_oneBos}) in the limit $m\rightarrow0$, as can be checked using the fact that ${}_3F_2\rightarrow 1$ when its argument vanishes. 

Finally, we remark that the CDD factor deformation of a free massive boson theory was considered in \cite{Dubovsky:2013ira}, where the first two orders of the perturbed action were constructed by a direct diagrammatic analysis, in perfect agreement with (\ref{eq:VLagrangian}), which generalizes this result to all orders and to a generic potential. 

\section{Exact cylinder  partition function}\label{subsec:PartFunc}
 In this Section we consider the $t$-deformation of rational CFTs and show how to compute the modified cylinder partition function. This can be  achieved, at least in principle, using  exact formulas for the perturbed Affleck and Ludwig ``ground-state degeneracy'' (commonly known as g-function) \cite{Affleck:1991tk, LeClair:1995uf, Dorey:1999cj, Dorey:2004xk,  Dorey:2009vg, Pozsgay:2010tv}. 
In the following, this method  is  applied to the theory of a free massless Majorana fermion: the Ising model CFT with  conformal boundary conditions.  The final analytic expression for the excited state g-functions is very simple, suggesting  that the form of the result  might  be  model independent. 
To support this conjecture, we will not work in the framework of the exact g-function formula extracted from \cite{Dorey:2009vg}, which would need to be generalized on a case-by-case basis. Instead, we shall adapt an alternative method originally devised by L\"uscher and  Weisz  \cite{Luscher:2004ib} for free massless bosons in the effective string theory context. This more general and powerful approach is implemented in Section \ref{sec:dualityOC}, leading to a confirmation of the Ising model result, which appears to be valid for a generic CFT. The construction also provides a compact integral expression for the cylinder partition function in terms of the CFT data. 

Let us start by setting the stage. The cylinder partition function of the theory, with boundary conditions $(\alpha, \beta)$ applied at the bottom and top of a cylinder of length $\LL$ and circumference $\rr$, can be written in two equivalent representations:
\begin{align}
\mathcal{Z}_{(\alpha,\beta)}(\LL,\rr) &= \sum_{n=0}^{\infty}e^{-\rr E^{(\alpha, \beta)}_{n}(\LL)} , &\text{(open channel)} \label{eq:open},\\
&= \sum_{n=0}^{\infty} g^{(\alpha)}_n(\rr) \, g^{(\beta)}_n(\rr) e^{-\LL E_n(\rr)},  &\text{(closed channel)} \label{eq:closed}.
\end{align}
The excited state g-function $g^{(\alpha)}_n(\rr)$ is defined by: 
\begin{equation}
 g^{(\alpha)}_n(\rr) = \langle \alpha|n \rangle  , 
\end{equation}
where $|n \rangle$ are the,  properly normalized, energy eigenvectors of the Hamiltonian defined on a ring with  circumference $\rr$.
 For the $t$-deformed theory, the closed channel energies are given by (\ref{eq:NGEP}), while the energies in the open channel take the generic form \cite{Caselle:2013dra}:
 \beqa
 E_n^{(\alpha, \beta)}(R, t ; c_{\text{eff}} ) = - \frac{\rr}{2 t} + \sqrt{ \frac{\rr^2}{4 t^2} +\frac{\pi}{t} \, \left( n-\frac{c_{\text{eff}}}{24} \right)  },
 \eeqa
where $c_{\text{eff}}$ depends on the conformal family of the state and the choice of boundary conditions translates into selection rules for the propagating states. In the following we will determine the form of the sub-leading coefficients $g_n^{(\alpha)}(R)$. 

\subsection{The Ising model CFT perturbed by $\TbT$} \label{subsec:g}
We shall consider the theory of  free massless Majorana fermions. Denoting with $p$ the momentum of the right mover and with $−q$ the momentum of the
left mover, the two-body scattering  amplitudes are
\begin{equation}
S_{\pm \pm }(p,q)= -1,\,\, S_{+-}(p,q)=S_{-+}(p,q)=S(p,q) = e^{-i 2 t p_+ p_-}= e^{i 2 t p q},
\end{equation}
where $p=p_+$ and $q=-p_-$ are the energies of right and left-moving pseudoparticles, respectively. 
The fundamental  boundary reflection factor is \cite{Caselle:2013dra}
\begin{equation}
\mathcal{R}_0(p) = e^{i t p^2},
\end{equation}
while the general boundary-perturbing reflection factor has the form 
\begin{equation}
\mathcal{R}_{\vec{\delta}}(p) =  \mathcal{R}_0(p)  \prod_{j=1}^{\infty} \left( \mathcal{R}_{\delta_j}^{2j-1}(p) \right)\,,\,\,\,\, 
\mathcal{R}_{\delta_j}^{2j-1}(p) = e^{(ip)^{2j-1}\delta_j}.
\label{RFfull}
\end{equation}
Each factor in the product appearing in the rhs of  (\ref{RFfull}) corresponds to a specific perturbation of the CFT boundary with a descendant of the identity operator, with coupling $\delta_j$. Let us first consider  the ground state  g-function. Following  \cite{Dorey:2009vg}, this quantity can be computed as:
\begin{equation}
\ln \left(g^{(\alpha)}_0(\rr,t) \right)= \ln \langle \alpha|0 \rangle^{\CFT} + \ln\left( {g_{0,\Sigma}}(\rr,t)\right) + \ln \left({g_0}(\rr,t)\right)  + 
\ln\left({g_{\vec{\delta}}}(\rr,t)\right) , \,\,
\end{equation}
with  $\alpha$ denoting one of the conformal boundary conditions $\alpha=(+),(-),(f)$ \cite{Dorey:2009vg}. The four terms contributing to $g^{(\alpha)}_0(\rr,t)$ are in general:
\begin{equation}
\ln \langle (+)|0 \rangle^{\CFT}=\ln  \langle (-)|0 \rangle^{\CFT}= -\ln \sqrt{2}, \,\,\,\ \ln \langle (f)|0 \rangle^{\CFT}=0,
\end{equation}
\begin{equation}
\ln ({g_{0,\Sigma}}(\rr,t)) = \sum_{j = 1}^{\infty} \frac{1}{2j - 1} \int_{\gamma_0} \frac{dp_1}{2\pi}...\int_{\gamma_0}\frac{dp_{2j - 1}}{2\pi}\frac{\varphi(p_1,p_2)}{1 + e^{\epsilon(p_1)}}...\frac{\varphi(p_{2j - 1},p_1)}{1 + e^{\epsilon(p_{2j - 1})}},
\end{equation}
\begin{equation}\label{gibi}
\ln({g_0}(\rr,t)) = \int_{\gamma_0}(\varphi_0(p) - \varphi(p,p)) {\cal L}(p)\frac{dp}{2\pi},
\end{equation} 
\begin{equation}
\ln({g_{\delta}}(\rr,t)) = \sum_j \ln{g_{\delta_j}}(\rr,t) = \sum_j \int_{\gamma_0} \varphi_{\delta _j}(p) {\cal L}(p) \frac {dp}{2\pi},
\end{equation}
where the integration contour is  $\gamma_0 = \R^+$
\bea\label{eq:kernels}
\varphi(p,q) &=& -i \partial_q \ln{S(p,q)}=2 t p\,, \,\,\,\varphi_0(p) = -i \partial_p \ln{\mathcal{R}_0(p)}=2 t p , \\
\varphi_{\delta_j}(p) &=&  -i \partial_p \ln{\mathcal{R}_{\delta_j}^{(2j - 1)}(p)} =(-1)^{j-1} (2j-1)\delta_jp^{2j-2}, 
\eea
${\cal L}(p) = \ln{(1 + e^{-\epsilon(p)})}$ and  the pseudoenergy $\epsilon(p)$ solves the (periodic) TBA equation. 
From (\ref{eq:kernels}), we see that $\ln \left(g_0(\rr,t) \right) = 0$, and
\begin{align}\label{sigma}
\begin{split}
\ln\left({g_{0,\Sigma}}(\rr,t) \right)&= \sum_{j = 1}^{\infty} \frac{1}{2j-1} \int_{\gamma_0}\frac{dp_1}{\pi} \frac{t \, p_1}{1 + e^{\epsilon(p_1)}} ... \int_{\gamma_0} \frac{dp_{2j-1}}{\pi} \frac{ t \, p_{2j-1}}{1 + e^{\epsilon(p_{2j-1})}}\\
&= \sum_{j=1}^{\infty}\frac{1}{2j-1} \left( \int_{\gamma_0} \frac{dp}{\pi} \frac{t \, p}{1 + e^{\epsilon(p)}} \right)^{2j-1} = \ln \sqrt{{\frac{1 + A_0}{1 - A_0}}},
\end{split}
\end{align}
where 
\begin{equation}
A_0 = \int_{\gamma_0} \frac{dp}{\pi} \frac{t \, p}{1 + e^{\epsilon(p)}}.
\end{equation}
To find $A_0$, we recall that the relevant  TBA equation is:
\begin{equation}
\epsilon(p) = \rr p +  p t \, E_0(\rr,t)
\quad \Rightarrow \quad
\partial_p \epsilon(p) = \rr + t \, E_0(\rr,t),
\end{equation}
\begin{equation}
E_0(\rr, t) = - \int_{\gamma_0} \frac{dp}{\pi} {\cal L}(p) = - \int_{\gamma_0} \frac{dp}{2\pi} \frac{2p}{1 + e^{\epsilon(p)}} 
\partial_p \epsilon(p).
\end{equation}
Therefore,
\begin{equation}\label{eq:A0}
E_0(\rr,t) = - \left(\rr + t E_0(\rr,t) \right)A_0/t \quad \Rightarrow \quad A_0 = -\frac{t \, E_0(\rr,t)}{\rr+ t \, E_0(\rr,t)}.
\end{equation}
Considering that the boundary conditions impose that only zero-momentum  states
contribute to $\mathcal{Z}_{(\alpha, \beta)}(\LL, \rr)$, the generalization to excited states is obtained by replacing  $\gamma_0 = \R^+$ 
with a  contour $\gamma_n$ encircling zeros of $(1+e^{\eps(p)}  )$ \cite{DoreyPerturbedCFT}. This means that $E_0 \rightarrow E_n$ in (\ref{eq:A0}). 
Going back to  (\ref{sigma}), the final result for the case $\vec{\delta}=0$ is 
\begin{equation}
g_n^{(\alpha)}(R, t) = \langle \alpha|n \rangle^{\CFT} \, \sqrt{\frac{ \rr}{ \rr +2 t \,  E_n(\rr,t)}}.
\label{eq:finalg}
\end{equation}
Equation (\ref{eq:finalg}) is the above mentioned simple formula for the deformed g-function. 
In the following section we shall prove that it remains valid  for more general CFTs. 
One interesting aspect of the formula is that the ``ground state degeneracy'' $g_0$ diverges precisely at the critical value of the radius $R_c$ already discussed in Section \ref{sec:Hagedorn}. This is a further manifestation of the Hagedorn (tachyon) transition in the model.  

Concerning the boundary corrections $(\ln{g_{\delta_i}})$, we find that the case $j=1$ is special:
\begin{equation}
\ln({g_{n,\delta_1}(\rr,t)}) = \delta_1 \int_{\gamma_n} \frac{dp}{2\pi}\,{\cal L}(p) = - \delta_1 \frac{E_n(\rr,t)}{2},
\end{equation}
as it amounts to a shift $\LL \rightarrow \LL+\delta_1$ (see also  \cite{Caselle:2013dra}):
\begin{equation}
\mathcal{Z}_{(\alpha,\beta)}(\LL,\rr,\delta_1) = \mathcal{Z}_{(\alpha,\beta)}(\LL+\delta_1,\rr, 0) .
\end{equation}
The associated boundary correction to the action $ {\cal A}$ is:
\begin{equation}
\delta {\cal A}_{\delta_1} = \fract{\delta_1}{2} \int_0^R dx \left( T_{yy}(x,0) + T_{yy}(x,\LL) \right).
\end{equation}
For general values of $j$, we have
\bea
\ln\left({g_{\delta_j}}(\rr,t)\right) &=& (-1)^{j-1}\delta_j(2j-1)\int_{\gamma_n}\frac{dp}{2\pi} p^{2j-2} \ln{\left(1 + e^{-(\rr + t \,E_n(\rr,t) )p}\right)} \nn \\
&=& (2j-1)(-1)^{j-1}\delta_j\left( \rr + t \, E_n(\rr,t) \right)^{1-2j}\int_{\gamma_n}\frac{dz}{2\pi}z^{2j-2}\ln{(1 + e^{-z})}  \nn \\
&=& \frac{(-1)^{j}}{2 \pi} \, \delta_j\left( \rr + t \, E_n(\rr,t) \right)^{1-2j} \, \Gamma(2j-1) \, {\text{Li}}^{(\gamma_n)}_{2j}(-1) . 
\label{poly}
\eea
In (\ref{poly}) a continuous branch of the polylogarithm, denoted  as ${\text{Li}}^{(\gamma)}_{m}(z)$, is involved.
\subsection{Cylinder partition function for more general perturbed CFT}\label{sec:dualityOC}
To treat a more general case, let us recall that, on a cylinder geometry, the set of Virasoro characters $\chi_i$ (corresponding to a conformal module labelled by $i$) of a rational CFT fulfil the following duality property (see, \cite{DiFrancesco:1997nk}):
\bea
\chi_i(q) &=& q^{-c/24} \text{Tr}_{i} \left(q^{L_0} \right)\,,\, \,\, q=e^{-2 \pi \rr/(2 \LL)}\, , \nn\\\
\chi_i(q) &=& \sum_j  S_{ij}\, \chi_j(\tilde{q})\,, \,\,\,  \tilde{q}=e^{-4 \pi \LL/\rr} .
\eea
This equation is directly related to the equivalence of the open/closed channel descriptions (\ref{eq:open}),(\ref{eq:closed}) of the CFT partition function:
\begin{align}
\begin{split}
\mathcal{Z}^{\CFT}_{\alpha \beta}(\LL,\rr) &= \text{Tr}(e^{-\rr \hat{H}_{(\alpha \beta)}(\LL)})=  \sum_{i} n^i_{\alpha \beta} \chi_i(q) = \sum_j \langle \alpha|j \rangle^{\CFT} \langle j|\beta \rangle^{\CFT} \chi_j(\tilde q) = \langle \alpha|e^{-\LL \hat{H}(\rr)} |\beta \rangle.
\label{Z1}
\end{split}
\end{align}
Adapting  an idea by L\"uscher and Weisz \cite{Luscher:2004ib}, we can recover  the $t$-deformed partition function as:
\bea
\mathcal{Z}_{\alpha\beta}(\LL,\rr) &=& \text{Tr}(e^{-\rr \hat{H}_{(\alpha \beta)}(\LL,t)}) \nn \\
&=& e^{-\frac{L R}{2 t} } \, \sum_{i} n^i_{\alpha\beta} \int_0^{\infty} \fract{dv}{\sqrt{4 \pi} v^{3/2}} e^{-\left(\fract{1}{4 v} + (\frac{\LL \rr}{2 t} )^2 v\right)}   
 \chi_i(e^{-\frac{\pi \rr^2 v }{t} }) \nn \\
&=&  e^{-\frac{L R}{2 t}} \, \sum_{j} \langle \alpha| j \rangle^{\CFT} \langle j |\beta \rangle^{\CFT} \int_0^{\infty} \fract{dv}{\sqrt{4 \pi} v^{3/2}}  e^{- \left(\fract{1}{4 v} + (\frac{L R}{2 t})^2 v \right)}    
 \chi_j(e^{-4 t  \pi/( \rr^2 v)}) \nn \\
&=& \sum_{j,n}   \langle \alpha|j \rangle^{\CFT} \langle j|\beta \rangle^{\CFT} \frac{\rr}{ \rr +2 t \,  E_{(j,n)}(\rr,t)} e^{-\LL E_{(j,n)}(\rr,t)}\nn \\
&=& \langle \alpha|e^{-\LL \hat{H}(\rr,t)} |\beta \rangle .
\label{Z11}
\eea
Therefore, 
\begin{equation}
g_{\alpha,(j,n)}(\rr,t)= \langle \alpha| j \rangle^{\text{CFT}} \, \sqrt{\frac{  \rr }{  \rr + 2 t \, E_{(j,n)}(\rr,t)}}.
\label{general}
\end{equation}
Equation (\ref{general}) generalises the Ising model result (\ref{eq:finalg}) to arbitrary CFTs, correspondingly equations (\ref{Z11}) provides a nice and compact integral representation for the perturbed partition function on a cylinder geometry. 
\section{Deformation of one-point correlation functions}
The simple deformation  rule (\ref{eq:invBurg}) for the energy spectrum leads to similarly simple changes of other physical quantities, such as one-point correlation functions on a cylinder.
In the following we shall implicitly assume that the $\TbT$ perturbation deforms the spectrum of a generic 2D quantum field theory as 
\begin{equation}
E_n(\rr,t)=E_n(\rr+ t \, E_n(\rr,t),0)\,,
\label{eq:mapping1}
\end{equation}
independently  from the integrability property of the original model. Consider a set $\{ \varphi_{i} \}$ of  scalar fields and the corresponding perturbed action on a cylinder:
\begin{equation}
{\cal A}_{(\vg, 0)} = \int_{cyl} \, dx  dy \, {\cal L}_{(\vg, 0)} \, =  \int_{cyl} \, dx  dy  \left(  {\cal L}_{\text{CFT}} + \sum_i  \, g_i \varphi_i (x,y)  \right).
\label{eq:PAction}
\end{equation}
One-point expectation values on this geometry can be obtained by differentiating the partition function $\mathcal{Z}(\LL, \rr, \vg)$ as
\beqa\label{eq:onepoint}
\langle \varphi_i(0,0) \rangle_{\rr}  = -\lim_{\LL \rightarrow \infty} \frac{1}{\LL \rr}  \partial_{g_i}\ln \mathcal{Z}(\LL, \rr, \vg) =
\frac{1}{\rr} \partial_{g_i} E_0(\rr, \vg),\,\,\,
\eeqa 
where $E_0(\rr, \vg)$ is the ground state energy on the ring with circumference $\rr$. Without loss of generality, we can choose a normalization of 
the spectrum consistent with the TBA/NLIE framework, such that  $E_0(\rr, \vg)$ vanishes in the $R \rightarrow \infty$ limit. One-point expectation values
on the plane are then encoded into the (anti) bulk coefficient ${\cal E}^{(\text{bulk})}(\vg)$ in $E_0(\rr, \vg)$  \cite{ZamolodchikovTBA}:  
\beqa
E_0(\rr, \vg)= E_0^{(\text{pert})}(\rr,\vg)- {\cal E}^{(\text{bulk})}(\vg,0) \,\rr ,
\eeqa
where, at least in principle,  the regular part $E_0^{(\text{pert})}(\rr,\vg)$ can be obtained using CFT perturbation theory. Therefore
\beqa
\langle \varphi_i(0,0) \rangle_{\rr}  =    \langle \varphi_i(0,0) \rangle_{\rr}^{(\text{pert})} - \langle \varphi_i(0,0) \rangle^{(\text{plane})}\,, \, \,\, 
\langle \varphi_i(0,0) \rangle^{(\text{plane})} =  \partial_{g_i} {\cal E}^{(\text{bulk})}(\vg,0)\,.
\eeqa
For integrable QFTs, a standard technique allows to  identify ${\cal E}^{(\text{bulk})}(\vg,0)$ by studying the $\rr \sim 0$ behaviour of TBA/NLIE equations \cite{ZamolodchikovTBA, KlassenMelzer}. 

For the current discussion, it is not necessary to set the couplings $\left\{g_i\right\}$ in (\ref{eq:onepoint}) to zero, so that the argument is valid also outside the (trivial) CFT 
limit, where all one-point functions would be zero on the cylinder. 
According to our assumptions, the $\TbT$ deformation affects the ground state energy according to (\ref{eq:mapping1}), and correspondingly one-point functions such as the ones 
defined in (\ref{eq:onepoint}) are modified as
\beqa
\langle \varphi_i(0,0) \rangle_{\rr} \rightarrow  \langle \varphi_i^{(t)}(0,0) \rangle_{\rr, t} &=&  \frac{1}{\rr} \,  \frac{1}{(1 - \left. t \, 
\partial_{\rr} E_0(\rr , \vg) \right|_{\rr=\widetilde{\rr}}) }   \left. \partial_{g_i} E_0(\rr, \vg )\right|_{\rr=\widetilde{\rr}} \, \nn \\
 &=&  \frac{\widetilde{\rr}}{\rr} \frac{  \langle  \varphi_i(0,0) \rangle_{ \widetilde{\rr}, t=0} }{(1- \left. t \, \partial_{\rr} E_0(\rr ,\vg) \right|_{\rr=\widetilde{\rr}}) },
\label{eq:def1p}
\eeqa
where $\varphi_i^{(t)}(x,y) $ is a redefined field,
\beqa
\varphi_i^{(t)}(x,y)   = \partial_{g_i} {\cal L}_{(\vg,t)}(x,y) ,
\eeqa
and  $\widetilde{\rr} = \rr + t \,  E_0(\widetilde{\rr},\vg)$. 
Notice that in (\ref{eq:def1p}) the denominator vanishes precisely at the critical points discussed in Section \ref{sec:Hagedorn}. Apart from the  universal factor in the denominator, the expectation values of the deformed fields $\varphi_i^{(t)}$  
are simply connected by a change $\rr \rightarrow \tilde \rr$ to the undeformed ones in the $t=0$ theory. This replacement governs both the spectrum and one-point functions, and it is tempting to imagine that general correlation functions could also be obtainable by an equally simple rule. However, even for the deformed free boson case, there are many points in the previous derivation which cannot be straightforwardly adapted to the calculation of multi-point correlators. Finally, identifying the $\mathcal{O}(R)$  bulk coefficient $\Epsilon^{(\text{bulk})}(\vg, t) $ from the small-$R$ behaviour of the ground state energy\footnote{An alternative way to obtain the same result is to note that, as shown in detail in \cite{FedorSasha}, the energy with bulk included $E_0^{(\text{pert})}(\rr, \vg)$ satisfies the same Burgers equation but for a simple change of variables. Notice that in the scheme used in \cite{FedorSasha} the mass scale depends on the deformation parameter, while throughout this paper we are keeping it fixed at its undeformed value.},
we also find:
\beqa
\langle \varphi_i^{(t)}(0,0) \rangle^{(\text{plane})}_t  = \frac{  \langle  \varphi_i(0,0) \rangle^{(\text{plane})}_{t=0} }{( 1 + t \, {\cal E}^{(\text{bulk})}(\vg, 0) )^2 }.
\eeqa 

The deformation of the fields, $\varphi_i \rightarrow \varphi_i^{(t)}$, appearing in (\ref{eq:def1p}) is a consequence of the nonlinearity of the definition of the $\TbT$ operator, which generates, through  (\ref{eq:lagrangian}), a nontrivial $\vg$-dependence in the Lagrangian ${\cal L}_{(\vg,t)}$. 
It is interesting to look at the explicit expression for the deformed fields $\varphi_i^{(t)}$ for the models discussed in Section \ref{sec:FreeBoson}. 

$\bullet$  \emph{Free massless bosons: }

\noindent Let us start by  considering the Lagrangian
\beqa
{\cal L}_{(g, 0)} =  \partial \phi \, \bar \partial \phi + g \, \varphi,
\eeqa
where $\varphi = f(\phi)$ is a generic field. We will apply the ideas of the previous section, setting $g \rightarrow 0$ at the end of the calculation to study the case of the pure $t$-deformation of a free massless boson. Therefore, we will compute $\left.  \varphi^{(t)}   = \partial_{g} {\cal L}_{(g,t)} \right|_{g=0}$. The deformed action  ${\cal L}_{(g,t)} $ can be simply obtained from (\ref{eq:VLagrangian}) with the replacement $m^2 V(\phi) \rightarrow g \, \varphi$. 
We find
\begin{eqnarray*}
\partial_{g}\mathcal{L}_{(g, t)}\Big\vert_{g=0} & = & \varphi +\frac{\varphi}{2\sqrt{\pi}}\sum_{j=1}^{\infty}(-4t)^{j}\frac{\Gamma\left(j+\frac{1}{2}\right)}{\Gamma\left(j+2\right)}\left(\partial\phi\bar{\partial}\phi\right)^{j}\frac{j^{2}-1}{2j-1}\\
 & = & \varphi-\frac{\varphi}{2}\left(1-\frac{1+2t\partial\phi\bar{\partial}\phi}{\sqrt{1+4t\partial\phi\bar{\partial}\phi}}\right).
\end{eqnarray*}
Comparing with (\ref{eq:tautheta}), the result above can be rewritten as 
\beqa\label{eq:NGtheta}
\partial_{g}\mathcal{L}_{(g, t)}\Big\vert_{g=0} = \varphi^{(t)} = \varphi \,[1 -t\,\theta ] ,
\eeqa
where the trace of the stress-energy tensor $\theta=\theta_0$ is defined in (\ref{eq:tautheta}).  
While the formulae are more intricate, we checked that the final result (\ref{eq:NGtheta}) holds for the case of an arbitrary number of free bosons, with $\theta=\theta_0$ defined in (\ref{eq:tauthetaN}).

$\bullet$ \emph{Single interacting boson: }

\noindent In this case, the starting point is the Lagrangian for a single bosonic field with a generic potential:
\beqa\label{eq:Lagrg}
{\cal L}_{(g, m, 0)} =  \partial \phi \, \bar \partial \phi + m^2 \, V(\phi)  + g \, \varphi ,
\eeqa
Setting ${\cal V} = m^2 V(\phi)$, in this case we find
\begin{eqnarray*}
\varphi^{(t)} &=& \partial_{g}\mathcal{L}_{(g,m, t)}\Big\vert_{g=0}  = \frac{\varphi}{\left(t {\cal V}  -1\right)^{2}}\\
&+&\frac{\varphi}{2\sqrt{\pi}}\sum_{j=1}^{\infty}(-4t)^{j}\frac{\Gamma\left(j+\frac{1}{2}\right)}{\Gamma\left(j+2\right)}\left(\partial\phi\bar{\partial}\phi\right)^{j}\frac{j^{2}-1}{2j-1}\;_{3}F_{2}\left(\left.\begin{array}{ccccc}
-j & , & \frac{3-j}{2} & , & 1-\frac{j}{2}\\
 & \frac{3}{2}-j & , & 1-j
\end{array}\right|-\frac{{\cal V} }{\partial\phi\bar{\partial}\phi}\right)\;.\label{eq:caseV}
\end{eqnarray*}
By a term by term inspection we found that the result can be rewritten in a very simple form:
\beqa
\varphi^{(t)}  =\varphi \, \left[1-2t\theta-t^{2}\left(\tau\bar{\tau}\right)\right]\;,
\eeqa
where the trace of the stress-energy tensor $\theta$ and the composite field $\tau \bar \tau $ are defined in the standard way in terms of the Lagrangian (\ref{eq:Lagrg}). 
Expression (\ref{eq:caseV}) reduces to the previous result (\ref{eq:NGtheta}) when ${\cal V}=0$, thanks to equation (\ref{eq:thetattb}).
\section{Conclusions}\label{Sec:Conc}
The study of 2D quantum  field theories  perturbed by irrelevant operators is still a largely unexplored research topic. These perturbations may lead to singular RG flows where the UV fixed point is not well-defined. An important arena where such flows appear prominently is the study of effective string theories, such as the ones used to describe confining flux tubes in non-Abelian gauge theories. 
Recently, some of the powerful tools associated to integrable models have made an appearance in this context, suggesting that exact results can be obtained for some of these irrelevant deformations. In this paper we investigated some general exact properties of a class of such flows related to the $\TbT$ composite operator. Among the infinite number of possible  perturbations of a given QFT, the latter operator displays very special and universal features.

From the point of view of a Lagrangian description, this  deformation is  formally defined through 
(\ref{eq:lagrangian}),(\ref{eq:TTbar}). These equations  were deduced using integrable models tools, such as the TBA and the NLIE, together with Zamolodchikov's formula in \cite{Zamolodchikov:2004ce}, but can be considered as a definition of a special way of deforming an arbitrary QFT, irrespective of integrability. Moreover, under such a deformation, the energy levels evolve according to the simple hydrodynamic equation (\ref{eq:inhomBurg}), where the deformation parameter $t$ enters as a formal time variable. This, in turn, allows  to reconstruct the perturbed spectrum from the unperturbed energy levels through the simple non-linear  mapping (\ref{eq:inSol}). For zero momentum states, in particular, (\ref{eq:inSol}) reduces to
\begin{equation}
E_n(R,t)=E_n(R+ t \, E_n(R,t),0)\,.
\label{mapping}
\end{equation}
Therefore, while the study of generic irrelevant perturbations in QFT by means of perturbative and non-perturbative methods remains  very  problematic, thanks to the exact mapping  (\ref{eq:inSol}), (\ref{mapping}),  this perturbation appears to be  surprisingly easy to treat. This  fact may  open the way to the study of  conformal field theories perturbed by  $\TbT$ plus  arbitrary, non necessarily integrable,  combinations of  relevant operators using the standard Truncated Conformal Space Approach (TCSA) \cite{Yurov:1989yu} method or its  RG-improved variant \cite{Giokas:2011ix}. 
We feel that research on this topic may  clarify important aspects concerning the appearance of singularities in effective QFT, and hopefully be useful in the effective string framework. It is well known that the flux tube spectrum is very well approximated, at relative large inter-quark separation, by the formula  (\ref{eq:NGEP}) or by its open-string analogous,  but deviations are  observed at intermediate length scales (see, for example  \cite{Brandt:2016xsp} for a nice  up-to-date compact review). Particularly interesting is the idea that also massive excitations could propagate on the flux-tube. 
It would be wonderful if TCSA could help to find a precise perturbed CFT interpretation of the pseudoscalar particle whose presence on the 4D QCD flux tube was recently conjectured on the basis of numerical data and TBA-inspired computations \cite{Dubovsky:2013gi}, or the massive mode discovered through Monte Carlo simulations in the 3D $U(1)$ gauge-theory \cite{Caselle:2014eka}.

It is important to stress that  (\ref{eq:NGEP}) was originally found by a light-cone-gauge quantization of the string, and that there is no known way to obtain this spectrum from a direct quantization of the static gauge action in a generic target space dimension $D \geq 3$. Furthermore, the light-cone spectrum is known to be incompatible with  Lorentz invariance in the D-dimensional target space, except for the cases $D=3$ and $D=26$. In particular, a universal deviation from the spectrum (\ref{eq:NGEP}) is observed at order $\mathcal{O}(1/R^5)$ (for the excited states), when the string is quantized preserving target-space Lorentz invariance \cite{Aharony:2010db} (see also \cite{Aharony:2013ipa},\cite{dubovsky2012effective},\cite{Brandt:2016xsp} for reviews). For this reason, we find remarkable that within the current setup,  a direct correspondence between the spectrum (\ref{eq:NGEP}) and the Nambu-Goto action written in static gauge is established in an  arbitrary dimension\footnote{A related but not totally equivalent result shows that, in $D=3$ and $D=26$, the CDD factor (\ref{eq:genCDD}) corresponds to the correct Lorentz-invariant quantization of the Nambu-Goto action \cite{Dubovsky:2015zey}. Further, the latter analysis shows that, without the inclusion of extra worldsheet degrees of freedom, integrability is incompatible with target space Lorentz symmetry in $D \neq 3$, $26$. }. 
A possible explanation is suggested by the fact that perturbative calculations in \cite{Aharony:2013ipa} using the static-gauge action with a continuum, strictly two-dimensional regularization scheme show that the light-cone spectrum is reproduced at least up to -- and including -- order $\mathcal{O}(1/R^5)$. The approach taken in this paper is indeed strictly two-dimensional, suggesting that, within the same class of schemes, the result is valid at all orders\footnote{We thank Marco Meineri for suggesting this 
interpretation. }.

Coming back to the problem of target space Lorentz invariance, from the point of view of the current setup one may try to enforce this symmetry by adding further irrelevant operators, associated to corrections of the S-matrix, on top of the $\TbT$-related ones considered in this paper. The form of the extra phase shift proposed in \cite{dubovsky2015flux} for the $D=4$ case seems to suggest that the first additional perturbing operator is related to odd-spin symmetry generators, perhaps associated to some extended symmetry such as the $W_3$-
algebra of the three states Potts model \cite{Zamolodchikov:1987zf}. 
Despite obvious complications, it would be nice to study this issue and understand if the current approach can at least partially be modified to better approximate Lorentz-invariant effective string models. 

 %
\appendix
\medskip
\noindent{\bf Acknowledgments --}
We would like to thank Michele Caselle, Beatrice Conti, Davide Fioravanti, Ferdinando Gliozzi, Simone Piscaglia for useful discussions, and especially Sergei Dubovsky, Victor Gorbenko, Marco Meineri and Davide Vadacchino for clarifying discussions on issues related to effective string theories.  We are also very grateful to Fedor Smirnov and Sasha Zamolodchikov for comments and encouragement at various points, and for sharing with us a copy of their paper ``On space of integrable quantum field theories'' \cite{FedorSasha} before publication.

 The research leading to these results has received funding from the People Programme (Marie Curie Actions)
of the European Union’s Seventh Framework Programme FP7/2007- 2013/ under REA Grant
Agreement No 317089 (GATIS), and was partially supported by INFN grant FTECP and the UniTo-SanPaolo research  grant Nr TO-Call3-2012-0088 {\it ``Modern Applications of String Theory'' (MAST)}. 
The work of SN was supported by the European Research Council (Programme “Ideas” ERC- 2012-AdG 320769 AdS-CFT-solvable). 
\bibliographystyle{JHEP2}
\bibliography{Biblio} 

\end{document}